\documentclass[runningheads,a4paper]{llncs}

\usepackage{csquotes} %\enquote{}

\usepackage[T1]{fontenc}
\usepackage{lmodern}
\usepackage{microtype}
\usepackage{hyperref}
\usepackage[all]{hypcap}
\usepackage{amssymb}
\usepackage{graphicx}

\usepackage{subfigure}
\usepackage{floatflt}

\usepackage{color,soul}

\usepackage{listings}

\usepackage{pgfplots,wrapfig}

\usepackage{url}
\urldef{\mailsa}\path|{daniel.ritter,manuel.holzeitner}@sap.com|    
\newcommand{\keywords}[1]{\par\addvspace\baselineskip
	\noindent\keywordname\enspace\ignorespaces#1}

\definecolor{mygray}{rgb}{0.8,0.8,0.8}

\newcommand{\eg}{e.\,g., }
\newcommand{\ie}{i.\,e., }

\begin{document}

\mainmatter  % start of an individual contribution

\title{Qualitative Analysis of Integration Adapter Modeling}

\titlerunning{Qualitative Analysis of Integration Adapter Modeling}

\author{Daniel Ritter \and Manuel Holzleitner}

\institute{Technology Development, SAP SE\\
	Dietmar-Hopp-Allee 16, 69190 Walldorf, Germany\\
	\mailsa}


\toctitle{Lecture Notes in Computer Science}
\tocauthor{Authors' Instructions}
\maketitle
\begin{abstract} \emph{Integration Adapters} are a fundamental part of an integration system, since they provide (business) applications access to its messaging channel. However, their modeling and configuration remain under-represented. In previous work, the integration control and data flow syntax and semantics have been expressed in the Business Process Model and Notation (BPMN) as a semantic model for message-based integration, while adapter and the related quality of service modeling were left for further studies.
	
%Through the rise of new, more data-intensive messaging scenarios (\eg, online player position tracking in sports management, internet of things), the conventional adapters (\eg, HTTP, FTP) are now complemented by data-centric software- (\eg, Extract/Transform/Load, Complex Event Processing) and novel hardware-solutions, which brings new challenges to their modeling.
	
In this work we specify common adapter capabilities and derive general modeling patterns, for which we define a compliant representation in BPMN. The patterns extend previous work %of \emph{Integration Flow} modeling 
by the adapter flow, evaluated syntactically and semantically for common adapter characteristics. 
	
\end{abstract}

\section{Introduction}
Although Enterprise Application Integration (EAI) continues to receive widespread focus by organizations offering them as means of integrating their conventional business applications with each other, with the growing amount of cloud applications and with their partners' systems, the integration adapter modeling is currently under-represented.

In this document we summarize a brief quantitative analysis of the currently used integration adapter types and their tasks according to the classification in \cite{DBLP:conf/caise/RitterHolzleitner15}. The analysis is conducted on the widely used, open source integration system \emph{Apache Camel} \cite{apachecamel2015,apacheCamel13}. Based on the quantitative analysis, we conducted qualitative studies with integration experts in design thinking work shops and surveys. The studies target the modeling aspects of integration adapters from \cite{DBLP:conf/caise/RitterHolzleitner15}.

\section{Apache Camel Component Analysis} \label{sec:analysis}
\begin{figure}[!ht]
	\begin{center}$
		\begin{array}{cc}
		\subfigure[Polling Consumers]{\label{fig:stat_polling}\includegraphics[width=0.5\textwidth]{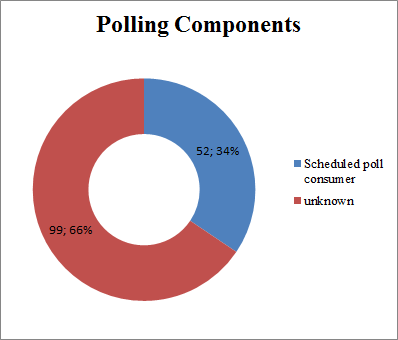}}
		\subfigure[Consumer vs. Producer ratio]{\label{fig:stat_general}\includegraphics[width=0.5\textwidth]{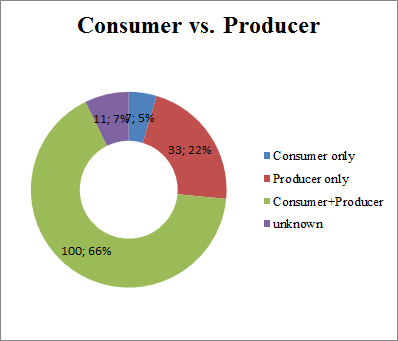}} &
		\end{array}$
	\end{center}
	\caption{Message Endpoint Analysis of 119 Apache Camel Component bundles with in total 151 Components. The term \enquote{Component} is used equivalent to \enquote{Adapter}.}
	\label{fig:statistics}
\end{figure}

To get a basic overview of existing integration components we implemented a system to introspect all Apache Camel \cite{apachecamel2015,apacheCamel13} component bundles in the \texttt{org.apache.camel} group in version $2.13$, that were registered in the \emph{Central Maven Repository} as of September 2014. With the help of byte code analysis, by using the open-source library \emph{ASM}\footnote{OW2 Consortium, visited 02/2015: \url{http://asm.ow2.org/}}, we automatically extracted basic capabilities from all of the $151$ found adapters (cf. Listing \ref{lst:all-components}). As such, we checked whether the consumer of an adapter extends from an scheduled poll consumer class to find out components which do not provide event-based consumers (cf. Listing \ref{lst:scheduled-poll-consumer-components}, Figure \ref{fig:stat_polling}). We categorized the components whether they support producers, consumers or both (cf. Listings \ref{lst:consumer-only}, \ref{lst:producer-only}, \ref{lst:consumer-producer}, Figure \ref{fig:stat_general}), by checking if the component provides the necessary implementation classes for the \emph{Camel Producer} / \emph{Consumer} interface. Please note, that some components provide producer implementations that in fact consume messages rather than send messages, \eg the \texttt{pop3} component can only be used to consume / poll for e-mails, although it provides an implementation for the producer interface. Such producer implementations are commonly used to compute a poll for messages, triggered by an event-message (event-based polling). This sub-categorization and the discovery of some, other capabilities could not be extracted automatically via byte code analysis, because this would require quite complex data-flow analysis, and remains to be done in the future.

\lstset{ %
  backgroundcolor=\color{mygray},
  breakautoindent=false,
  breakindent=0em}
\begin{lstlisting}[language=java,caption={All Components (151)},label={lst:all-components},breaklines=true,basicstyle=\ttfamily\scriptsize]
ahc, ahc-ws, ahc-wss, amqp, apns, atmosphere-websocket, atom, avro, aws-cw, aws-ddb, aws-s3, aws-sdb, aws-ses, aws-sns, aws-sqs, aws-swf, bean-validator, box, cache, cmis, cometd, cometds, context, couchdb, crypto, cxf, cxfbean, cxfrs, disruptor, disruptor-vm, dns, dropbox, ejb, elasticsearch, exec, facebook, flatpack, fop, freemarker, ftp, ftps, gauth, geocoder, ghttp, ghttps, glogin, gmail, google-drive, gora, gtask, guava-eventbus, hazelcast, hbase, hdfs, hdfs2, http, http4, https, https4, ibatis, imap, imaps, infinispan, irc, ircs, javaspace, jclouds, jcr, jdbc, jetty, jgroups, jms, jmx, jpa, jt400, kafka, kestrel, krati, ldap, lpr, lucene, metrics, mina, mina2, mongodb, mqtt, msv, mustache, mvel, mybatis, nagios, netty, netty-http, netty4, netty4-http, nntp, openshift, optaplanner, pop3, pop3s, quartz, quartz2, quickfix, rabbitmq, restlet, rmi, rnc, rng, routebox, rss, salesforce, sap-netweaver, schematron, scp, servlet, sftp, sip, sjms, smpp, smpps, smtp, smtps, snmp, solr, solrCloud, solrs, somp, spark-rest, splunk, spring-batch, spring-event, spring-integration, spring-ldap, spring-redis, spring-ws, sql, ssh, stax, stream, string-template, twitter, velocity, vertx, weather, websocket, xmlrpc, xmlsecurity, xmpp, xquery, yammer, zookeeper
\end{lstlisting}

\begin{lstlisting}[language=java,caption={Scheduled Poll Consumers components (52)},label={lst:scheduled-poll-consumer-components},breaklines=true,breakautoindent=false,basicstyle=\ttfamily\scriptsize]
apns, atom, aws-s3, aws-sqs, bean-validator, cmis, cxfbean, dropbox, ejb, facebook, freemarker, ftp, ftps, gora, hbase, hdfs, hdfs2, http, http4, https, https4, ibatis, imap, imaps, jclouds, jpa, jt400, krati, msv, mustache, mvel, mybatis, nntp, openshift, optaplanner, pop3, pop3s, rnc, rng, sftp, smtp, smtps, splunk, sql, ssh, stax, string-template, twitter, velocity, weather, xquery, yammer
\end{lstlisting}

\begin{lstlisting}[language=java,caption={Consumer only components (10)},label={lst:consumer-only},breaklines=true,breakautoindent=false,basicstyle=\ttfamily\scriptsize]
atom, hazelcast, jetty, jmx, quartz, quartz2, rss, servlet, snmp, spark-rest
\end{lstlisting}

\begin{lstlisting}[language=java,caption={Producer only components (33)},label={lst:producer-only},breaklines=true,breakautoindent=false,basicstyle=\ttfamily\scriptsize]
ahc, aws-cw, aws-ddb, aws-sdb, aws-ses, aws-sns, crypto, dns, elasticsearch, exec, fop, gauth, geocoder, glogin, gmail, jclouds, jdbc, jt400, ldap, lpr, lucene, metrics, nagios, sap-netweaver, schematron, scp, solr, solrCloud, solrs, spring-batch, spring-ldap, xmlrpc,xmlsecurity
\end{lstlisting}

\begin{lstlisting}[language=java,caption={Consumer \& Producer components (108)},label={lst:consumer-producer},breaklines=true,breakautoindent=false,basicstyle=\ttfamily\scriptsize]
ahc-ws, ahc-wss, amqp, apns, atmosphere-websocket, avro, aws-s3, aws-sqs, aws-swf, bean-validator, box, cache, cmis, cometd, cometds, context, couchdb, cxf, cxfbean, cxfrs, disruptor, disruptor-vm, dropbox, ejb, facebook, flatpack, freemarker, ftp, ftps, ghttp, ghttps, google-drive, gora, gtask, guava-eventbus, hbase, hdfs, hdfs2, http, http4, https, https4, ibatis, imap, imaps, infinispan, irc, ircs, javaspace, jcr, jgroups, jms, jpa, kafka, kestrel, krati, mina, mina2, mongodb, mqtt, msv, mustache, mvel, mybatis, netty, netty-http, netty4, netty4-http, nntp, openshift, optaplanner, pop3, pop3s, quickfix, rabbitmq, restlet, rmi, rnc, rng, routebox, salesforce, sftp, sip, sjms, smpp, smpps, smtp, smtps, splunk, spring-event, spring-integration, spring-redis, spring-ws, sql, ssh, stax, stomp, stream, string-template, twitter, velocity, vertx, weather, websocket, xmpp, xquery, yammer, zookeeper
\end{lstlisting}

%\section{Background Information on Adapter Workshop} \label{sec:workshop}
%- add figures of background\\
%- number of participants \\
%- etc\\

\section{Qualitative Analysis of the Adapter Modeling Approach} \label{sec:result}
The adapter characteristics and the modeling approach were part of a survey-based analysis with $20$ integration experts. The surveys were partially conducted as interviews (due to expert availability). The results are briefly discussed subsequently and have been anonymized, where necessary.

The survey was set up mostly with free-text fields and multiple-choice sections, the questions and the diagrams from the paper \cite{DBLP:conf/caise/RitterHolzleitner15}. Not all participants answered all questions.

\subsection{Adapter Characteristics}
\begin{figure}[!ht]
	\begin{center}$
		\begin{array}{cc}
		\subfigure[Adapter Types]{\label{fig:survey_adapter_types}\includegraphics[width=0.45\textwidth]{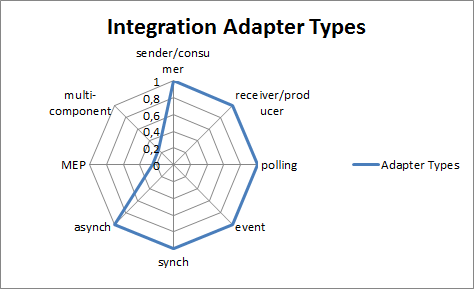}}
		\subfigure[Adapter Tasks]{\label{fig:survey_adapter_tasks}\includegraphics[width=0.45\textwidth]{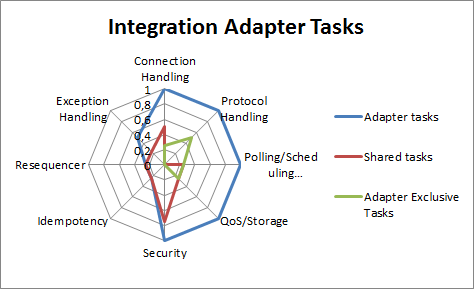}} &
		\end{array}$
	\end{center}
	\caption{Adapter Types and Tasks (100\% equals to 1.0)}
	\label{fig:statistics}
\end{figure}

Some of the adapter characterisitcs have been already analyzed experimentally in Section \ref{sec:analysis}. However, we cross-checked some aspects (\ie adapter types and tasks) in the survey as introductory questions to gain some more information on the participant's background. Figure \ref{fig:survey_adapter_types} shows that all participants know the basic adapter types and their separation into polling or event-based consumers and producers, as well as synchronous (synch) and asynchronous (asynch) communication. Surprisingly, not all survey participants differentiate between synch/asynch communication and the \emph{Message Exchange Patterns} (MEPs) \texttt{inOnly} / \texttt{inOut} (cf. \cite{DBLP:conf/caise/RitterHolzleitner15}). Same applies to the Apache Camel multi-component modeling, which allows to specify multiple transport protocols in one Camel component.

For the adapter tasks, all participants named (technical) connection and protocol handling, scheduling, storage and Quality of Service (QoS) support as well as transport-level security aspects like encryption, authentication. Figure \ref{fig:survey_adapter_tasks} shows the summarized responses. Notably, most of the participants did not name the resequencer and idempotency characteristic separately, but saw them included in the QoS support. Others correctly named idempotency as task of the receiver, which is considered the only save case. However, if a receiving backend system does not support that, the integration system (\eg producer adapter) has to take over. The exception handling was mentioned by little less than half of the participants, which might show a lack of awareness for the topic and could require an additional educational offering.

The exception handling, resequencer, and security (\ie message-level: en-decrypt, signing, not communication channel security) were mostly seen as shared tasks between the adapters and integration pipelines. Topics that were seen exclusively important for the adapter are protocol / format handling.

\subsection{Adapter Flow}
The \emph{Adapter Flow} (AF) was considered a requirement for many scenarios by all participants. However, most participants argued for consumer AFs, due to the possibility of message pre-processing before handing the message to the integration process. Apart from some exceptions, the post-processing could be moved to the integration process. Prominent exceptions were seen in the idempotent message handling and potential flexibility to build scenario-specific extensions. The modularity and architectural aspects were only brought up by few participants. For them it is important to be able to run the AFs on non-integration runtime systems (\eg Extraxt/Transform/Load or Complex Event Processing-Tools for more data-centric processing).

\subsection{Bridges}
Most participants like the explicit modeling of \enquote{bridges}, however, have doubts about complexity: nobody wants to model these constructs piece by piece. Hence the experts proposed a pattern-based modeling approach similar to the \emph{Enterprise Integration Pattern} (EIP) modeling \cite{DBLP:journals/corr/Ritter14}, which could allow to drag\&drop the bridges from the palette. One problem that was mentioned is about multiple the placement of the same pattern and simultaneous changes afterwards. For that, the design time tool would have to offer refactorings for multiple patterns, at the same time, through \enquote{where-used} support, and capabilities that allow to capture an adapted pattern as user-defined, \enquote{re-usable pattern}.

The participants require such a modeling approach for scenario-based variations of the default bridge implementations. For simple adapters (\eg that do not require special pre- / post-processing or QoS support), the participants mostly argued for the BPMN \emph{Message Flow}-based modeling, which means \enquote{over-defining} the BPMN constructs and adding a property sheet.

\subsection{Security Aspects}
The explicit modeling of security relevant information like certificats/key stores, was controversially discussed. Partners and consultants require more configuration capabilities for certificate handling and an explicit association from the AFs. That means, a key store could be referenced from multiple integration processes.

On the other hand, more business-near experts argues that the explicit modeling will be too complex in most cases.

\subsection{Message Queuing}
For message queuing the participants stated that the diagrams are understandable as long as only two integration processes are connected via queues in one IFlow. Variants in which four integration processes were connected to one single messaging system or duplicated messaging systems in the same IFlow were used, were regarded as too complex. The transactional dequeuing variant was received well als it makes the transactional boundaries visible, however the modeling of the boundaries might not be required in all cases as they might be derived from the tasks themselves.

As an alternative, the messaging system modeled as a BPMN participant allows to show the inner workings of the messaging system, which was received as informative but too complex for some users to understand. On the one hand, technical and implementation-oriented users see the need for a flexible and comprehensive modeling of the queuing and the messaging system. On the other hand, business-oriented users would like to have an high-level overview on the overall integration by hiding technical details such as messaging and queueing in general (only noticeable as QoS). 

\begin{figure}[!ht]
	\centering
	\includegraphics[width=0.9\columnwidth]{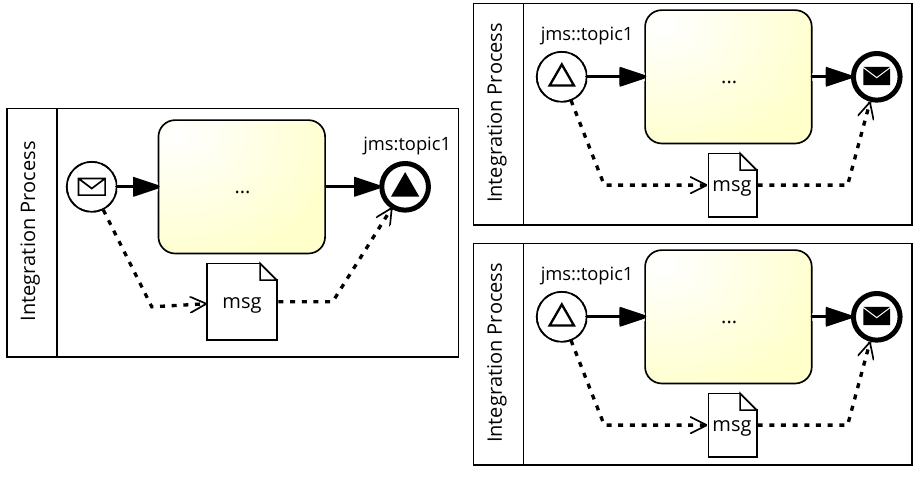}
	\caption{Alternative Topic Modeling using \emph{BPMN Signal End Event}.}
	\label{fig:topic2}
\end{figure}

All participants agreed that the modeling aproach via a BPMN data store is superior to the alternative modeling aproach via BPMN signals (cf. Topic Modeling in Figure \ref{fig:topic2}). Finding the matching names for the topic on one diagram without a visual hint via connectors reduces the overall understandability of an IFlow by making connectivity too implicit. 

\subsection{Reply Flow}
The reply flow was received well in the survey as it allows to model important post-processing steps in the IFlow. Also other use-cases were identified where the modeling of the reply flow is required. However, here the reply flow adds complexity to the overall IFlow, thus identifying the need for good tool support that allows for hiding / un-hiding the reply flow. 

\subsection{Cross-Tenant, Network Modeling}
For cross-tenant connectivity between IFlows the modeling via enqueu-/dequeuing on a messaging system it gets visible that they are communicating in an asynchronous fashion. Thus, the messaging system modeled as BPMN data store can function as the boundary to other IFlows (in other tenants) and can be used to split multiple diagrams to reduce the complexity. However, to get an high-level overview it must be possible to combine all connected diagrams to an high-level overview.

\subsection{Quality of Service Modeling}
In the survey some participants prefer to model QoS on sender adapter as a property without explicitly model redelivery, reqsequencing and de-duplication via idempotency repository. Other participants prefer the improved capabilities and flexibility of an integration system supporting detailed QoS modeling. However, it is important to note that some of the QoS tasks are preferably located either on the receiver or the sender adapter, \eg the redelivery should usually be done on sender side and the deduplication should be modeled in the receiver adapter.

\section{Conclusions}
The work relates to the classification of integration adapters and modeling approches described in \cite{DBLP:conf/caise/RitterHolzleitner15}. Based on a quantitative analysis (\ie, real-world case studies, integration expert interviews and surveys), this work gives a brief, but comprehensive overview of integration adapter types and tasks. With the qualitative user studies modeling alternatives are evaluated and user preferences captured (\eg the value of a modular, pattern-based and explicit modeling approach) as well as its downsides (\eg modeling complexity, technical diagrams). These identified aspects will be further analized in quantitative user studies and an extension of the modeling language (\eg conditional data flows).

%\renewcommand{\indexname}{List of Patterns}
%\printindex

\section*{Acknowledgments.} We thank all participants of the expert workshops, surveys and interviews as well as Volker Stiehl and Ivana Trickovic for valuable discussions on BPMN.

\bibliographystyle{abbrv}
\bibliography{analysis}

\end{document}